\documentclass[aps,prl,twocolumn,groupedaddress,showpacs]{revtex4}

\usepackage{graphicx}% Include figure files
\usepackage{dcolumn}% Align table columns on decimal point
\usepackage{bm}% bold math

\begin{document}

\preprint{APS/123-QED}

\title{Flow of He~II due to an Oscillating Grid in the Low Temperature Limit}

\author{H.A. Nichol, L. Skrbek$^{*}$, P.C. Hendry, and P.V.E. McClintock}
 %\altaffiliation[Also at ]{Department of Physics, Lancaster University, Lancaster LA1 4YB,
%UK.}%Lines break automatically or can be forced with \\
%\author{}
\affiliation{%
Department of Physics, Lancaster University, Lancaster LA1 4YB,
UK\\ $^{*}$Joint Low Temperature Laboratory, Institute of Physics
ASCR and Charles University, \\V
  Hole\v{s}ovi\v{c}k\'ach 2, 180\,00 Prague, Czech Republic}%

\date{\today}% It is always \today, today,

\begin{abstract}
The macroscopic flow properties of pure He~II are probed in the
limit of zero temperature using an oscillating grid. With
increasing oscillation amplitude the initially pure superflow
changes abruptly: the resonant frequency decreases and the
response becomes strongly nonlinear, attributable to the growth of
a boundary layer of quantized vortices that increases the
effective mass of the grid. On further increase of oscillation
amplitude, the flow undergoes a transition to turbulence.
\end{abstract}

\maketitle

%\section{\label{sec:level1}First-level heading:\protect\\ The line
%break was forced \lowercase{via} \textbackslash\textbackslash}

Although the exotic flow properties of He~II have been a subject
of intensive investigation ever since the discovery of
superfluidity, accumulating a vast amount of experimental data and
theoretical knowledge \cite{Tough,Newton}, many important features
still remain to be explained. In the limit of low velocity, He~II
flow is very well described in terms of Landau's two-fluid model.
On increasing the flow velocity beyond a certain threshold,
however, quantized vortices appear. Their presence couples the
originally independent normal and superfluid velocity fields via a
mutual friction. Above $\sim$1.2~K, where He~II contains an
appreciable proportion of normal fluid, numerous investigators
observed that on exceeding a suitably defined Reynolds number,
He~II flow acquires an increasingly classical character: (i) the
surface of a rotating bucket forms a nearly classical meniscus
\cite{Osborne:50}; (ii) flow past a microsphere displays both
laminar and turbulent drag \cite{Wilfried}, as well as (iii) a
drag crisis \cite{vSci}; (iv) the energy spectrum of turbulent
He~II involves an inertial range with a classical K41 Kolmogorov
roll-off exponent of $-5/3$ \cite {Tabeling}; (v) the decay of
both grid-generated and counterflow turbulence displays a
classical character \cite{Skr, Skr1, Vinen, QTReview}. Although
these features are seen over a wide range of normal
fluid/superfluid density ratios, it is impossible to exclude the
possibility that this classical-like behaviour is caused by the
presence of the viscous normal fluid. Following oscillating
experiments with a tiny sphere \cite{Wilfried} and a very thin
vibrating wire \cite{Wire}, yielding interesting results
attributable to single vortices, there is a clear call to study
the {\it macroscopic} flow properties of He~II in the $T
\rightarrow 0$ limit where normal fluid is (almost) absent and
pure superflow can be investigated systematically.

\begin{figure}
%h=here, t=top, b=bottom, p=separate figure page
\begin{center}
%\leavevmode
\includegraphics[angle=0,width=1.00\linewidth]{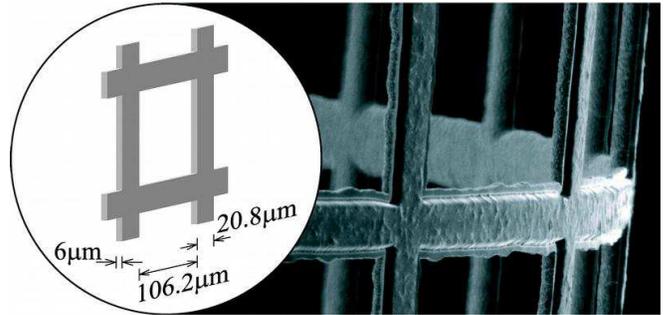}
\caption{Schematic drawing showing the geometry of the grid and an
electron microscope picture of the actual grid (sample cut off the
same sheet).}\label{grid}\end{center}\end{figure}

\begin{figure*}
%h=here, t=top, b=bottom, p=separate figure page
\begin{center}\leavevmode
\includegraphics[angle=90,width=.90\linewidth]{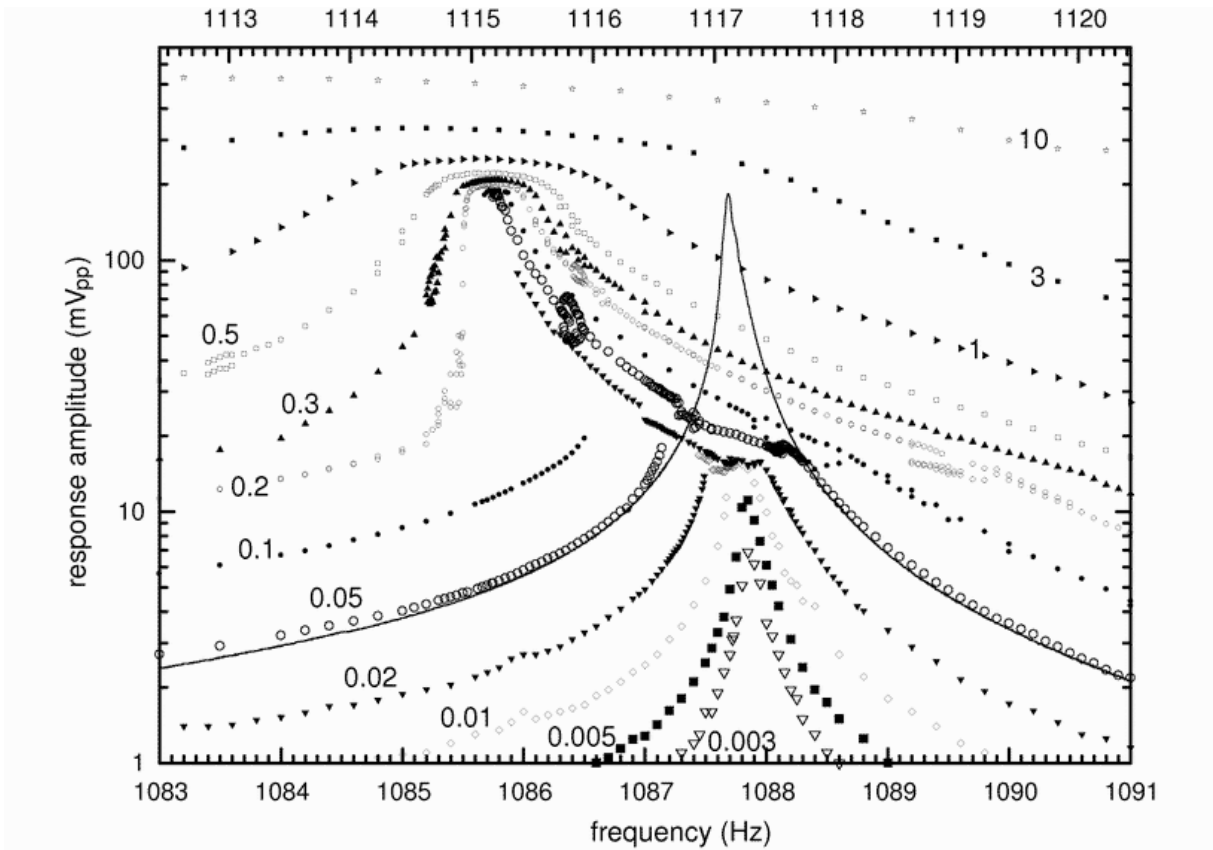}
\caption{Resonance curves measured at $10$ bar and nominally
$50$~mK using the memory oscilloscope for various drive levels (in
V$_{pp}$) as indicated. Doubled symbols in some places indicate
where stable beatings were observed while carefully sweeping the
drive frequency, typically in steps of 0.001 Hz. The solid line
represents the resonant response to a 0.05 V$_{pp}$ drive in
vacuum (measured using the lock-in amplifier, upper frequency
axes).} \label{resonances}
\end{center}\end{figure*}

Our tool for performing such investigations is the thin
electroformed nickel grid \cite{grid} of $70$\% transparency and
mesh size of $127$ $\mu$m shown schematically in Fig.\ \ref{grid}.
It constitutes a circular membrane $2R=8$ cm in diameter, tightly
stretched on a circular mild steel carrier, mounted horizontally,
equidistant from two gold-plated plane copper electrodes, each
drilled with 170 holes of 2~mm~ID to connect the grid space to the
rest of the experimental chamber. The electrodes are separated by
$d=2$ mm, immersed in a 1.5 liter volume of isotopically pure
$^4$He in an experimental cell attached to the mixing chamber of
the dilution refrigerator. The apparatus can be operated at all
pressures up to the solidification pressure. A static potential,
typically $V_{0}=500$ volts, is applied to the grid. A driving
potential $V_{1}=V_{10}\cos\omega t$ ($V_{10}$ $\ll V_{0}$)
applied to the upper electrode produces a driving force on the
grid of form $f_{d}=2 \varepsilon _{0} \varepsilon_{r} \pi\ R^{2}
VV_{1}/d^{2}$, where $\varepsilon_{0}$ and $\varepsilon_{r}$
denote respectively the relative permittivities of free space and
the liquid $^4$He. The grid can thus be considered as an
oscillating membrane \cite{Morrell} under uniform tension. In this
study we approximate its motion as one-dimensional and assume that
the oscillation amplitude is uniform across its area. Oscillations
of amplitude $\Delta D$ induce a signal of amplitude $V_{2}=V
\Delta D/(2d)$ on the lower electrode, which can be monitored with
a lock-in amplifier (which we use to investigate the low-drive
linear response of the grid) or directly with a memory
oscilloscope, allowing visualization of transient processes
resulting from the strongly nonlinear response at higher drives.
Allowing for a reduction in the induced voltage $V_{2}$ by a
factor of $(1+C_{c}/C)^{-1}\approx0.065$, where $C_{c}\cong 700$
pF is the capacitance of the connecting cable and $C\cong 47$ pF
is the capacitance between the grid and the lower copper
electrode, the response amplitude $|V_{2}|$ provides a direct
measure of the amplitude and spatially averaged peak velocity
$|v_{g}|=|\omega \Delta D|$ of the oscillating grid.

At the lowest oscillation amplitudes, the grid displays linear
behaviour in that its frequency response to the drive is a
Lorentzian of narrow width caused predominantly by nuisance
damping, with a quality $Q$ factor typically exceeding 5000. The
resonant frequency $f_{\rm res}$ can be altered temporarily by
increasing the pressure, which we attribute to quantized vortices
generated e.g.\ by the jet from the filling capillary: they
probably become pinned to inhomogeneities on the grid surface, and
between the grid and surrounding electrodes. Moving the grid
violently at high drive amplitude, however, is found to stabilize
$f_{\rm res}$ typically to within $\pm0.1$ Hz, presumably by
shaking off most of this pinned vorticity.

\begin{figure}
%h=here, t=top, b=bottom, p=separate figure page
\begin{center}\leavevmode
\includegraphics[angle=90, width=1.00\linewidth]{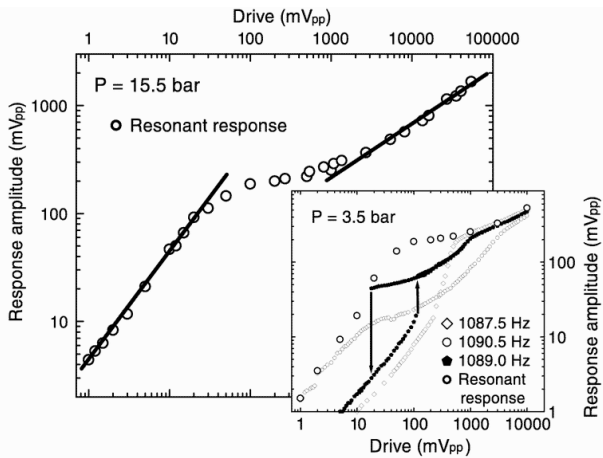}
\caption{Response amplitude of the grid versus the drive level at
resonance (main figure) and at fixed frequencies (inset). For
explanation of hysteresis, see text. The full lines indicate the
linear and square-root responses. Note that the main figure and
inset refer to different pressures.}\label{ampdrive}
\end{center}\end{figure}

Fig.\ \ref{resonances} shows our central experimental observation,
which for the ``cleaned" grid has been repeated for several
pressures spanning 0.3 $\leq p \leq$ 24.8 bar without appreciable
change. As the drive increases and the grid amplitude reaches the
first threshold (typically 10--20 mV$_{pp}$, corresponding to a
mean grid velocity of $0.3<v_{g}^{(1)}<0.6$ cm/s), the oscillation
amplitude at resonance continues to rise in proportion to the
drive (Fig.\ \ref{ampdrive}), but the resonant frequency suddenly
starts decreasing and the resonance curves acquire highly
nonlinear features. If one follows e.g.\ the resonance curve for
the 0.05 V$_{pp}$ drive down from 1091 Hz (Fig.\ \ref{resonances})
by slowly sweeping the drive frequency (in practice, digitally, in
steps of 0.001 Hz), the system displays the usual nearly
Lorentzian stationary response while the amplitude remains below
the first threshold. On further decreasing the drive frequency,
the response amplitude increases {\it above} this first threshold
until, shortly below 1086 Hz the amplitude suddenly collapses down
to a lower stable branch. On increasing the drive frequency again
from this point the system stays on the lower branch until about
1087.15 Hz, where a transition to the stable upper branch occurs.
These hysteretic loops are robust to temperature increase, at
least up to our maximum of  130 mK.

Within the hysteretic parameter range, beat phenomena are
occasionally seen between two apparently stable amplitude levels,
with envelopes of period $\sim$1~s (see, eg., Fig.\
\ref{resonances}, drive 0.05 V$pp$ around 1086.5 Hz); once
established, they are stable on the scale of hours. Small changes
of frequency do not kill them, but modify the upper and lower
amplitude levels between which beating occurs. Only with further
change of the driving frequency does the beating disappear. On
restoring the original frequency the response is not completely
reproducible; within some small frequency range beating might not
appear at all; and sometimes it reappeared only later, or not at
all. Despite considerable effort, we did not succeed in
establishing any fully repeatable pattern or well-defined
conditions for the appearance of beats \cite{beating}.

\begin{figure}
%h=here, t=top, b=bottom, p=separate figure page
\begin{center}
%\leavevmode
\includegraphics[angle=0,width=0.8\linewidth]{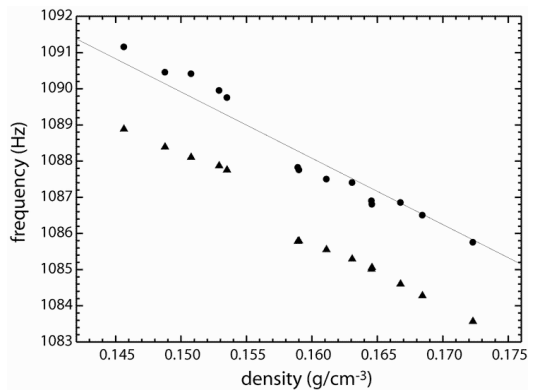}
\caption{Frequency as a function of He~II density at low drive
levels (circles) and for the second critical threshold (triangles)
The straight line extrapolates through the zero-density (vacuum)
resonant frequency of $f_0$=1117.2 Hz.}\label{freqshift}
\end{center}
\end{figure}

The fall in frequency with increasing drive (Fig.\
\ref{resonances}) reaches typically 2 Hz for all investigated
pressures (see Fig.~\ref{freqshift}), ceasing at a second
threshold amplitude (typically $200$ mV$_{pp}$ corresponding to a
mean grid velocity of about $v_{g}^{(2)}\approx6$ cm/s).
Observation of the two well-defined resonant frequencies, at all
investigated pressures shifted by $\approx2$~Hz, is a remarkable
feature of the superflow that, to our knowledge, has not
previously been reported. With further increase in drive, the
oscillation amplitude at resonance initially remains almost
constant, while the widths of the resonance curves increase
rapidly (see Fig.\ \ref{resonances}). Only for drive levels
exceeding by about an order of magnitude that for the second
threshold does the amplitude at resonance grow again; this time
approximately in proportion to the square-root of the drive, as
shown in Fig.\ \ref{ampdrive}. In this high-drive regime the
linewidth increases rapidly while the resonant frequency decreases
gradually, qualitatively in the manner expected for increasing
damping. Pronounced beatings were never observed in this
high-drive regime. If the oscillation amplitude is measured as a
function of drive level at fixed frequency, one observes clear
hysteresis within the frequency range containing the two stable
branches of the grid response (Fig.\ \ref{ampdrive}). Outside this
range, the drive dependences are nonlinear, but without any
hysteresis.

Measurements of the grid response at low temperature in vacuum
(smooth curve in Fig.\ \ref{resonances}) demonstrate that the
presence of the He~II has two effects. First, it down-shifts the
low-drive resonant frequency by about 30 Hz from its vacuum value
of $f_{0}= (1117.2\pm 0.05)$ Hz, depending on the pressure $p$ in
the cell (see Fig.\ \ref{freqshift}). This classical effect arises
from hydrodynamic enhancement of the mass $M$ of the grid by
$\Delta M_{H}= \beta \varrho_{He}(p) M/\varrho_{G}$, \cite
{grid,Brooks}. The measured $f_{0}$ and low drive resonant
frequencies yield $\beta=3.01\pm0.05$. Secondly, it is indeed the
He~II that is responsible for the observed nonlinearities
\cite{vacuum}.

We believe that the behaviour of the oscillating grid ought to be
understandable in terms of quantized vortices generated in He~II
by its motion. As one possibility, we may speculate that, on
exceeding the first threshold, a ``boundary layer" of vortex loops
(perhaps of horseshoe form \cite{kus,Samuels}) grows on the grid,
thus increasing its effective mass. Correspondingly, the resonant
frequency shifts down and the resonance curves acquire strongly
nonlinear features. Provided the vortex loops remain pinned and
that they do not reconnect to create free vorticity, the mass
enhancement would be non-dissipative, as observed (see below). We
note that small mass enhancements were observed earlier for a
sphere vibrating in He~II above 1~K \cite{hotsphere}, but without
associated critical thresholds.

It is interesting to characterize the first threshold by a
superfluid Reynolds number $Re_{s}=U_{\rm max}G / \kappa$, where
$U_{\rm max}$ stands for a peak critical flow velocity through the
grid window \cite{vmax}, $G$ is its linear size and $\kappa$
denotes the circulation quantum. Observed values of $Re_{s} \sim
10$ compare well with the critical $Re_{s}=UD/\kappa \approx 20 $
($U$ is the transport velocity and $D$ the diameter of the pipe)
found as a temperature independent threshold in pipe flow of He~II
at much higher temperature when the flow of the normal component
was inhibited by superleaks at both ends of the pipe
\cite{ToughPipe}. The generation of quantized vortex loops on the
surface of the grid by macroscopic superflow around it probably
involves growth from remanent vortex lines \cite{Awschalom}, or
from a ``plasma" of half vortex rings \cite{kus}. This idea is
supported by the following observation. While measuring the
response dependence on increasing the drive amplitude at the
resonant frequency, the system encounters a nucleation problem
when passing the first threshold: on some occasions the resonance
response stopped growing with increasing the drive level, and
jumped on the response/drive curve (Fig.\ \ref{ampdrive}) only
later. With decreasing drives this feature disappears, and the
response remains proportional to the drive level.

The frequency down-shift $\Delta f (p) =f_{\rm res}-f_{\rm
sh}\sim$ 2 Hz between the two critical thresholds might be
considered in terms of a boundary layer of thickness $\lambda$
that enhances the hydrodynamic effective mass of the grid by
$\Delta M_{g }^{\lambda} = A \lambda \varrho_{He}(p)$, where $A$
denotes the surface area of the grid. Requiring that the downward
shift of the resonance frequency corresponds to those observed
experimentally leads to
\begin{displaymath}
\lambda= {M+\Delta M_{g }^{\lambda}\over {A \varrho_{He}}}\left(
{{f_{\bf res}^{2}}\over{f_{sh}^{2}}}-1 \right)= \sqrt{{{2\nu_{\bf
eff}}\over\omega}}=0.53\pm0.05~\mu m
\end{displaymath}
which would correspond to an effective kinematic viscosity
$\nu_{\bf eff}\approx 10^{-5}$ cm$^2$/s. Extending the analogy to
classical fluids, we may estimate the expected resonant linewidth
due to the drag of such a hypothetical viscous fluid. A
straightforward calculation \cite{LL} based on estimating the flow
velocity gradient in the direction inside the fluid by $\langle
v_{g} \rangle /\lambda$ leads to a linewidth $\sim$1 Hz,
consistent with resonances for (e.g.\ Fig.\ \ref{resonances}, 0.2
and 0.5 V$_{pp}$ data) near the second critical threshold.
However, this picture seems unable to account for the re-entrant
character of the nonlinear resonances because the resultant
viscous broadening would tend to smear out the effect in question.

We infer that, beyond the second critical threshold (Fig.\
\ref{resonances}), the vortex loops start to reconnect. Free
vorticity is then shed into the liquid, corresponding to
dissipation of the vibrational energy and leading to the observed
increase in linewidth. Under these flow conditions, He~II behaves
in close analogy with a classical Navier-Stokes fluid: the
square-root behaviour (Fig.\ \ref{ampdrive}) of the resonant
response as a function of drive amplitude in this range is typical
of classical turbulent drag scaling, and it is therefore likely
that this threshold marks the onset of turbulence. After leaving
the grid, the quantized vorticity probably ``evaporates''
\cite{3HeBarenghi} as recently proposed in connection with
turbulence in superfluid $^3$He-B \cite{Fisher}.

The particular challenge posed by these results is to develop a
quantitative description of vortex dynamics in the macroscopic
flow around the moving grid, showing how it can give rise to the
observed amplitude-dependent mass enhancements and re-entrant
resonance curves, without dissipation. But the problem also
carries wider significance, and will repay careful study, because
the dynamical processes in question are probably fundamental to
the generation of quantum turbulence.

%The transition to turbulence occurs from the quasi-laminar regime
%of He~II flow, which already contains a certain amount of quantized
%vorticity. We believe that this quasi-laminar flow regime (when He
%II most likely mimics the viscous shear flow by appropriate
%arrangement of quantized vortex lines as is well known to take
%place in a rotating bucket) can possibly be identified with the
%turbulent state I using the classification scheme of Tough
%\cite{Tough}. The inferred onset of turbulence at the second
%threshold should thus correspond to the transition to Tough's
%turbulent state II.

We are grateful to N.J.\ Fullwood for making the electron
micrograph in Fig.\ \ref{grid}, and to D.I. Bradley, A. M.
Gu\'{e}nault, R.P.\ Haley, M.\ Krusius, D.G.\ Luchinsky, G.R.\
Pickett, P.\ Skyba and W.F.\ Vinen for fruitful discussions. The
research was supported by the Engineering and Physical Sciences
Research Council (UK) and the Czech Grant Agency under grant
GA\v{C}R 202/02/0251.


\begin{thebibliography}{99}
\bibitem{vmax} $U_{\rm max}$ is enhanced by a factor $\approx1.43$
due to the $70\%$ grid transparency; it decreases with increasing
pressure.
\end{thebibliography}
\end{document}